\documentclass[11pt,english]{paper}
\usepackage[T1]{fontenc}
\usepackage[latin9]{inputenc}
\usepackage{geometry}
\usepackage{float}
\usepackage{amsmath}
\usepackage{amsthm}
\usepackage{graphicx}
\usepackage{esint}

\makeatletter
\let\paperclassexample\example

\let\example\relax
\AtBeginDocument{%
  \@ifundefined{example}{\let\example\paperclassexample}{}
}
\numberwithin{equation}{section}
\numberwithin{figure}{section}

\makeatother

\usepackage{babel}
\begin{document}
\title{A Numerical Procedure for the Determination of the Pursuit Curve of
Objects with Uniformly Accelerated Motion.}
\author{Luis Rozas, luis.rozas@ulagos.cl{\huge\textbf{\date{}}}}
\institution{{\small Universidad de los Lagos, Departamento de Ciencias de la Ingenieria,
Puerto Montt, Chile}}
\maketitle
\begin{abstract}
In this article the pursuit problem of objects that moves with different
accelerations and initial speeds is studied. Initially, the situation
in which the escaping object moves in a straight line is considered.
Under this condition, and if both objects starts from rest, it is
shown that the trajectory of the chasing object match those obtain
for the case of uniform motions. When both objects have different
initial speeds and accelerations, we first note that if the escaping
object starts its motion with a speed which is function of the acceleration
and initial speed of the chasing object, then the pursuit curve is
the same obtained for no initial speeds. Latter solution is used to
solve the chase problem when both objects have different accelerations
and initial speeds. Finally, making use of the preceding results,
a numerical procedure is proposed for obtaining the pursuit curve
when the escaping object moves in an arbitrary path. 
\end{abstract}

\section{Introduction}

The chases and escapes problems are perhaps among the problems that
most have attracted attention from the mathematical community, going
back, as far as we know, to the famous paradox of the Greek philosopher
Zeno of Elea about the race between Achilles and the tortoise. Through
time different versions of pursuit problems were proposed. In 1638
Francis Godwin\textquoteright s published the story \textsl{The Man
in the Moone} in which an astronaut harnesses a wedge of 25 swans,
and flies to the moon. The swans fly at a constant speed and always
head toward the moon, in accordance with their annual migration, their
trajectory is then not a straight line. According to Godwin the time
needed to flight to the moon is twelve days, the return trip, which
is traveled following a straight line takes only eight days. Later
in 1732, the French mathematician, geophysicist, geodesist, and astronomer
Pierre Bouguer \cite{Bouguer1732} published an article in the \textsl{Memoires
de l'Academie Royale des Sciences} in which analyze the pursuit of
a pirate ship of a merchant vessel. 

Several chases and escapes problems have been solved since then. In
1982 A.A Azamov \cite{Azamov1982} consider the problem of a pursuer
with velocity equal to one, and an escaping object with bounded velocity
greater than one, giving for these conditions the escape strategy
that guarantees a positive constant lower bound for the distance between
the objects. Cyclic pursuit of n objects (bugs) that chase each other
in cyclic order, each moving at unit speed, was studied by T.J. Richardson
\cite{Richardson2001}. The pursuit-evasion problem of objects with
maximum speeds was studied by A.S. Kuchkarov \cite{Kuchkarov2010}
providing the necessary and sufficient conditions to complete the
capture of the evader. C. Hoenselaers \cite{Hoenselaers1995}, and
more recently Azevedo and Pelluso \cite{Azevedo2022}, tackle the
problem of the pursuit of objects with relativistic speeds, that is
taking into account that information propagates at a finite speed.
Other several chases and escapes problems can be found in the book
of Paul J. Nahin \cite{Nahin2007}, which contains an exhaustive collection
of pursuit problems and its solutions.

On the other hand the pursuit and evasion of several objects have
also brought the interest of many researchers, F.L. Chernous'ko \cite{Chernousko1976}
shown that if the speed of the escaping object is bounded can, by
remaining in a-neighborhood of a given motion, avoid an exact contact
with any number of pursuing objects whose velocities are less than
the velocity of the escaping one. Optimal strategies for the escaping
object when chased by many pursuers were also developed in the works
\cite{Ibragimov2005,Ibragimov2006,G.I.Ibragimov2015,Borowko1985}

The pursuit-evasion differential game of infinitely many inertial
players, meaning they moves with constant accelerations, and with
integral constraints on control functions, was studied by G.I Ibragimov
and M. Salimi \cite{Ibragimov2009}, showing that under certain conditions,
the value of the game, and the optimal strategies of players can be
found. 

The aim of this article is to study the pursuit problem of two objects
that moves with uniformly accelerated motions. To do this, the problem
is first analyzed in three cases, all of which correspond to situations
in which the escaping object moves in a straight line. Using the results
of these analyses, a numerical procedure in order to find the trajectory
of the pursuit object on the condition that the escaping object moves
on an arbitrary path is proposed.

\section{Case I: Both objects with uniform acceleration and zero initial speed}

As shown in figure \ref{fig:Case1}, let we consider that the pursuit
object begins its motion at the origin of the coordinate system, and
the escaping object begins at point $\left(d,b\right)$ and moves
along a straight line parallel to the Y axis. Also, let be $a_{p}$
and $a_{e}$ the acceleration of the pursuit and the escaping object
respectively. The condition of the problem dictates that the velocity
of the pursuit object always points to the escaping one. Considering
the figure \ref{fig:Case1}, this latter condition can be written
as:
\begin{figure}
\centering\includegraphics[scale=0.48]{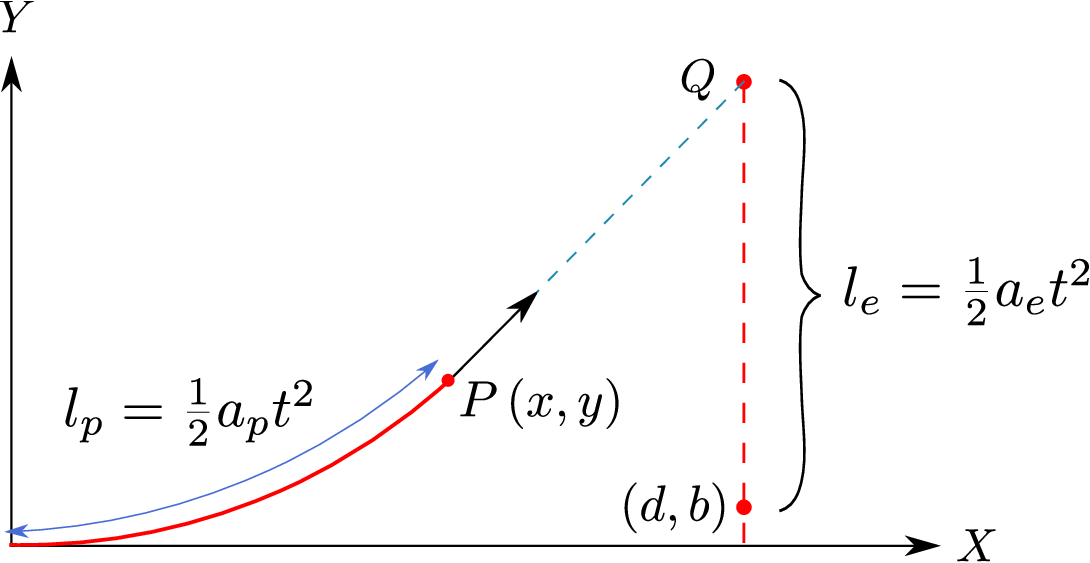}

\caption{Both objects have different accelerations but zero initial speeds.
After an arbitrary amount of time the pursuit object is located in
point $P\left(x,y\right)$ and points to the escaping object located
in $Q$. The velocity vector of the pursuit object, always points
towards $Q$. \protect\label{fig:Case1}}

\end{figure}
\begin{equation}
y^{'}=\frac{\frac{1}{2}a_{e}t^{2}+b-y}{d-x}\label{eq:Acc_cte}
\end{equation}
where $t$ is the time. The distance traveled by the escaping object
can be expressed in terms of the distance traveled by the pursuit
one simple as $l_{e}=\tfrac{a_{e}}{a_{p}}l_{p}$. On the other hand
this latter distance can be expressed as $l_{p}=\intop_{0}^{x}\sqrt{1+y^{'{}^{2}}}\,dx$.
Using these results, the equation \ref{eq:Acc_cte} can be rewritten
as:

\begin{equation}
y^{'}\left(d-x\right)-b+y=\frac{a_{e}}{a_{p}}\intop_{0}^{x}\sqrt{1+y^{'{}^{2}}}\,dx\label{eq:Acc_cte_1a}
\end{equation}

After taking derivatives to both sides of the equation \ref{eq:Acc_cte_1a}
and rearrange the terms we obtain the following ordinary differential
equation for the trajectory of the pursuit object:

\begin{equation}
y^{''}=\frac{a_{e}}{a_{p}}\cdot\frac{\sqrt{1+y^{'{}^{2}}}}{d-x}\label{eq:Acc_cte_1}
\end{equation}

This equation is similar to those obtained by P. Ptak and J. Tkadlec
\cite{Ptak1996}, except by the fact that in their case the objects
has no acceleration instead has constants speeds. The equation \ref{eq:Acc_cte_1}
can be reduced in one order after putting $u\left(x\right)=y^{'}\left(x\right)$.
The initial conditions of the problem are $y\left(0\right)=0$ and
$y^{'}\left(0\right)=b/d$, thus the solution of the differential
equation \ref{eq:Acc_cte_1} can be written in closed form as:

\begin{equation}
\frac{y\left(x\right)}{d}=\frac{1}{2}\left(\frac{\sqrt{\rho^{2}+1}-\rho}{1+\alpha_{a}}\left(1-\frac{x}{d}\right)^{1+\alpha_{a}}-\frac{\sqrt{\rho^{2}+1}+\rho}{1-\alpha_{a}}\left(1-\frac{x}{d}\right)^{1-\alpha_{a}}\right)+\frac{\rho+\alpha_{a}\sqrt{\rho^{2}+1}}{1-\alpha_{a}^{2}}\label{eq:TrayCaseI}
\end{equation}
where $\alpha_{a}=a_{e}/a_{p}$, $\rho=b/d$. It is interesting to
note that when $a_{e}=0$, meaning $\alpha_{a}=0$ the equation \ref{eq:TrayCaseI}
reduce simply to the straight line $y=\rho x$, which correspond to
the case in which the escaping object remains still. The capture,
if occurs, will happen when $x=d$, in which case we have:
\begin{equation}
y\left(d\right)=d\left(\frac{\rho+\alpha_{a}\sqrt{\rho^{2}+1}}{1-\alpha_{a}^{2}}\right)\label{eq:yCapture}
\end{equation}

If $\alpha_{a}<1$ then the pursuit object reach the escaping one.
The equation \ref{eq:yCapture} can be used to found the total distance
traveled by the escaping object, $L_{T}=y\left(d\right)-b$, till
being caught as follows:
\begin{equation}
\frac{L_{T}}{d}=\frac{\rho+\alpha_{a}\sqrt{\rho^{2}+1}}{1-\alpha_{a}^{2}}-\rho\label{eq:dt_Case1}
\end{equation}

\begin{figure}
\centering\includegraphics[scale=0.8]{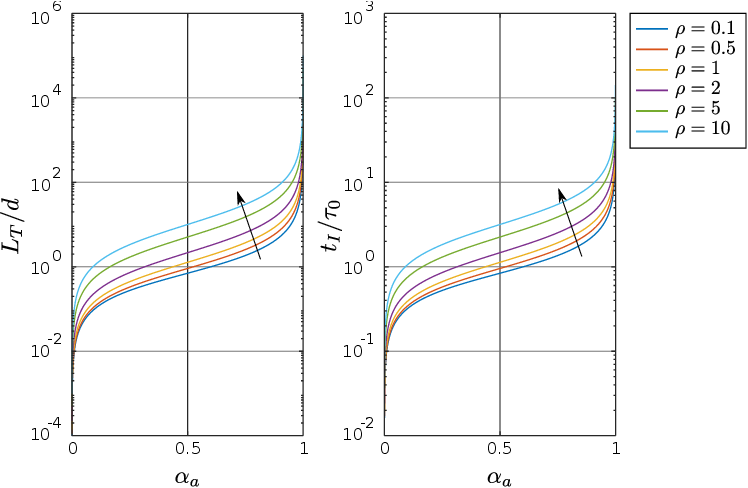}

\caption{Graphical representation of the distance traveled by the escaping
object relative to $d$, left figure, and the normalized total time
of persecution, right figure. Arrows indicate the increasing direction
of $\rho$\protect\label{fig:GraphCaseI}}
\end{figure}

The total time of the persecution for this case, $t_{I}$, can be
calculated recalling that the movement of the escaping object occurs
under uniform acceleration with zero initial speed:

\begin{equation}
\frac{t_{I}}{\tau_{0}}=\sqrt{\frac{\rho+\alpha_{a}\sqrt{\rho^{2}+1}}{1-\alpha_{a}^{2}}-\rho}\label{eq:NormTimeI}
\end{equation}
where $\tau_{0}=\sqrt{2d/a_{e}}$ is the time needed to travel the
distance $d$ from rest and with constant acceleration equal to $a_{e}$.
From the equation \ref{eq:NormTimeI} the time of capture when $a_{e}=0$
is equal to: $t_{I}=\sqrt{2\sqrt{b^{2}+d^{2}}/a_{p}}$, that is the
time that the pursuit object needs to travel in a straight line from
the origin to point $\left(d,b\right)$. In the figure \ref{fig:GraphCaseI}
the graphical representations of equations \ref{eq:dt_Case1} and
\ref{eq:NormTimeI} are shown.

\section{Case II: Both objects with uniform acceleration, escaping object
with initial speed equal to $\upsilon_{po}a_{e}/a_{p}$ }

Before analyze the general case, when both objects have different
initial speeds and different accelerations, let us consider the situation
depicted in figure \ref{fig:FigCase2} in which the initial velocity
of the escaping object is equal to: $\upsilon^{\ast}=\upsilon_{p0}a_{e}/a_{p}$,
where $\upsilon_{p0}$ is the initial velocity of the pursuit object.
In this case the distance traveled by the pursuit object is equal
to: $l_{p}=\frac{1}{2}a_{p}t^{2}+\upsilon_{p0}t$, and the distance
traveled by escaping one is: $l_{e}=\frac{1}{2}a_{e}t^{2}+\upsilon^{\ast}t$.
As in the previous case these both distances are related as: $l_{e}=\frac{a_{e}}{a_{p}}l_{p}$.
Therefore the slope of the trajectory of the pursuit object can be
written as:
\begin{figure}
\centering\includegraphics[scale=0.45]{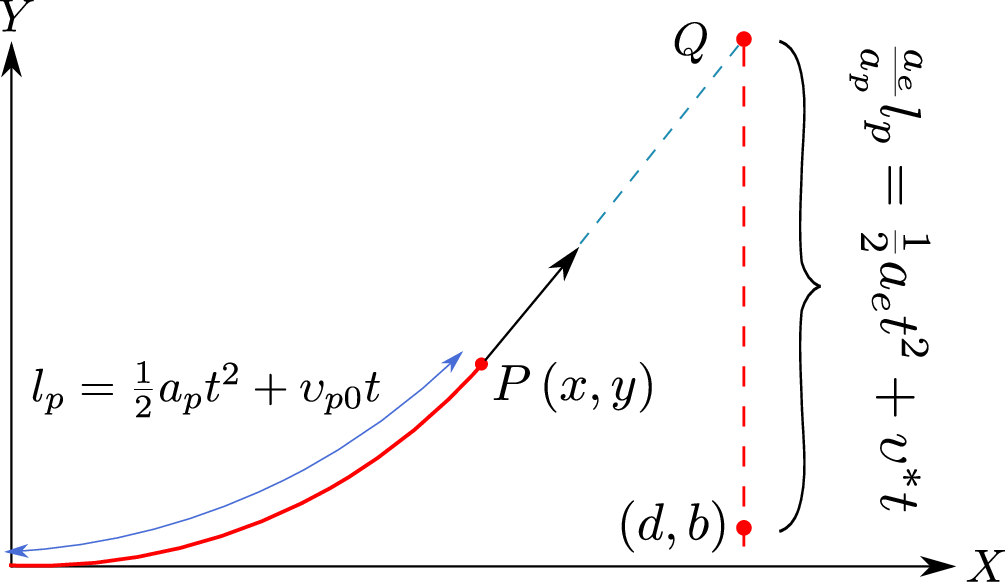}

\caption{The initial speed of the pursuit object is equal to $\upsilon_{p0}$,
and the initial speed of the escaping one is $\upsilon^{\ast}=\upsilon_{po}a_{e}/a_{p}$.
The pursuit object is located in point $P\left(x,y\right)$ and its
velocity always points to the escaping object located in $Q$. \protect\label{fig:FigCase2}}
\end{figure}
\begin{equation}
y'=\frac{\frac{a_{e}l_{p}}{a_{p}}+b-y}{d-x}\label{eq:Vinitial}
\end{equation}

After rearranging terms and remembering that $l_{p}=\intop_{0}^{x}\sqrt{1+y^{'{}^{2}}}\,dx$
we obtain:
\begin{equation}
y^{''}=\frac{a_{e}}{a_{p}}\cdot\frac{\sqrt{1+y^{'{}^{2}}}}{d-x}\label{eq:Case2-diffeq}
\end{equation}
which is the same differential equation of the case with no initial
speed for both objects, equation \ref{eq:Acc_cte_1}. Thus the trajectory
of the pursuit object is equal to the previous case. The difference
here regards with the persecution time. Using the equation \ref{eq:dt_Case1}
the total time of the chase in this case, $t_{II}$, can be obtained
from: $\frac{1}{2}a_{e}t_{II}^{2}+\upsilon^{\ast}t_{II}=L_{T}$, which
leads to:
\begin{equation}
t_{II}=-\frac{\upsilon^{\ast}}{a_{e}}+\sqrt{\frac{\upsilon^{\ast2}}{a_{e}^{2}}+\frac{2L_{T}}{a_{e}}}
\end{equation}

The latter expression can be rewritten in dimensionless way, naming
$\upsilon_{0}=a_{e}\tau_{0}$, and $\alpha_{\upsilon0}=\upsilon_{p0}/\upsilon_{0}$

\begin{equation}
\frac{t_{II}}{\tau_{0}}=-\alpha_{a}\alpha_{\upsilon0}+\sqrt{\alpha_{a}^{2}\alpha_{\upsilon0}^{2}+\frac{\rho+\alpha_{a}\sqrt{\rho^{2}+1}}{1-\alpha_{a}^{2}}-\rho}\label{eq:Case2_time}
\end{equation}

\begin{figure}
\centering\includegraphics[scale=0.7]{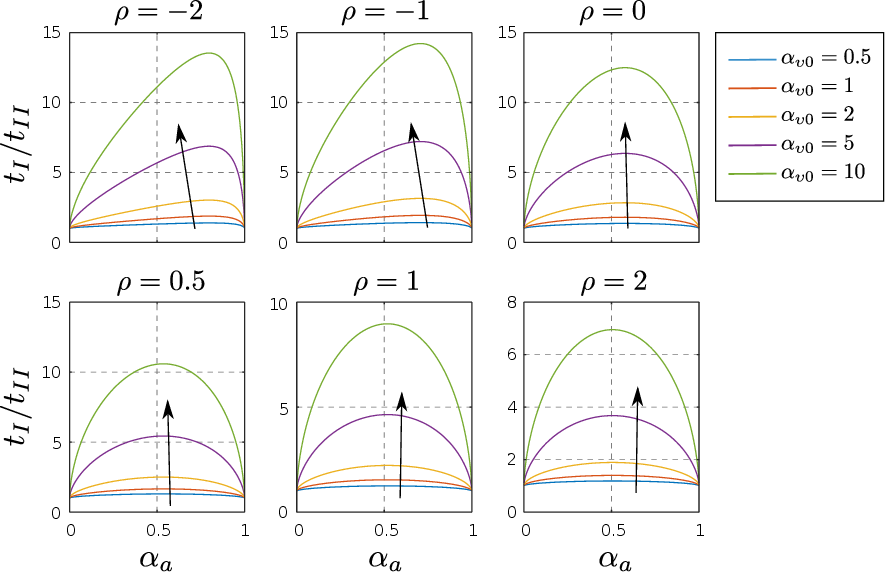}

\caption{Comparison between the chasing time for case i relative to chasing
time for case ii for different $\alpha_{a}$, $\alpha_{\upsilon0}$
and $\rho$ values. Arrows indicate the increasing direction of $\alpha_{\upsilon0}$.
\protect\label{fig:NormTime_Case2}}
\end{figure}

In the figure \ref{fig:NormTime_Case2} the quotient $t_{II}/t_{I}$
is plotted against $\alpha_{a}$ for various values of $\rho$ and
$\alpha_{\upsilon0}$. When $\alpha_{a}=0$ the escaping object just
remains stationary, the chasing time obtained from equation \ref{eq:Case2_time}
is in this case equal to:
\begin{equation}
t_{II}=-\frac{\upsilon_{p0}}{a_{p}}+\sqrt{\frac{\upsilon_{p0}^{2}}{a_{p}^{2}}+\frac{2\sqrt{b^{2}+d^{2}}}{a_{p}}}
\end{equation}
which coincide with the time needed by the pursuit object to travel
in straight line from the origin to point $\left(d,b\right)$ with
acceleration $a_{p}$ and initial velocity $\upsilon_{p0}$. 

\section{Case iii: Different accelerations and initial speeds }

Lets consider now the case when both objects have different accelerations
and initial speeds, being $\upsilon_{e0}$ the initial speed of the
escaping object. In order to find the trajectory of the pursuit object
the results of the previous case will be used here. Referring to the
figure \ref{fig:Case3}, the slope of the path of the pursuit object
can be written as:
\begin{figure}
\centering\includegraphics[scale=0.5]{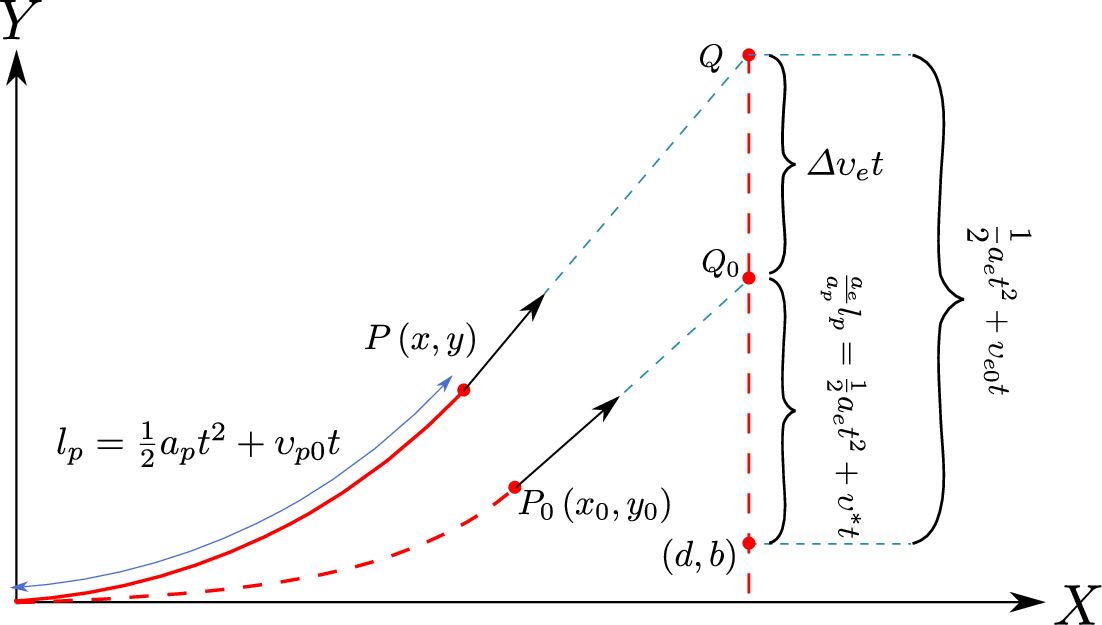}

\caption{The case when both objects have different accelerations and initial
speeds. Point $P$ represent the position of the pursuit object in
a given instant of time, and the point $P_{0}$is the corresponding
position of the pursuit object when the initial velocity of the escaping
object is $\upsilon^{\ast}$, ie for case ii scenario. The solution
of the case ii will be used in order to find the path of the pursuit
object when both objects have different accelerations and initial
speeds. \protect\label{fig:Case3}}
\end{figure}

\begin{equation}
y'=\frac{\frac{1}{2}a_{e}t^{2}+\upsilon_{e0}t+b-y}{d-x}\label{eq:Case3_Bruto}
\end{equation}
which, according to figure \ref{fig:Case3} is equivalent to:

\begin{equation}
y'=\frac{\frac{a_{e}l_{p}}{a_{p}}+\varDelta\upsilon_{e}t+b-y}{d-x}=\frac{\frac{a_{e}l_{p}}{a_{p}}+b-y_{0}+\varDelta\upsilon_{e}t+y_{0}-y}{d-x}\label{eq:Case3}
\end{equation}
where $\varDelta\upsilon_{e}=\upsilon_{e0}-\upsilon^{\ast}$. Naming
$\left(x_{0},y_{0}\right)$ the position of the pursuit object when
the initial velocity of the escaping one is $\upsilon^{\ast}$ and
using the equation \ref{eq:Vinitial} we can write:
\begin{equation}
y'\left(d-x\right)=y_{0}'\left(d-x_{0}\right)+\varDelta\upsilon_{e}t+y_{0}-y\label{eq:Case3Mod}
\end{equation}

The persecution time is related to the path length of the pursuit
object as: $t=\tfrac{1}{a_{p}}\left(\sqrt{\upsilon_{p0}^{2}+2a_{p}l_{p}}-\upsilon_{p0}\right)$,
replacing in the equation \ref{eq:Case3Mod} and after rearrange terms
we obtain:
\begin{equation}
y'\left(d-x\right)-y_{0}'\left(d-x_{0}\right)-\left(y_{0}-y\right)+\frac{\varDelta\upsilon_{e}\cdot\upsilon_{p0}}{a_{p}}=\frac{\varDelta\upsilon_{e}}{a_{p}}\sqrt{\upsilon_{p0}^{2}+2a_{p}l_{p}}\label{eq:diffCase3}
\end{equation}

Squaring and taking derivative to the latter equation, and remembering
that $l_{p}=\intop_{0}^{x}\sqrt{1+y^{'{}^{2}}}\,dx$, leads us to
the following second order differential equation for the trajectory
of the pursuit object:
\begin{equation}
y''=\frac{\varDelta\upsilon_{e}^{2}\sqrt{1+y^{'{}^{2}}}}{a_{p}\left(d-x\right)\cdot A}+y_{0}''\left(\frac{d-x_{0}}{d-x}\right)\label{eq:Case3_EDO}
\end{equation}
where $A$ is the left side of the equation \ref{eq:diffCase3}. Equation
\ref{eq:Case3_EDO} provides the trajectory of the pursuit object,
which can be obtained numerically using the solution of case ii as
a starting point. If instead of using equation \ref{eq:Case3_EDO}
we attempt to obtain the trajectory of the pursuit object directly
dealing with the equation \ref{eq:Case3_Bruto}, that is without using
the solution of case ii, we should replace the time of persecution
in the expression for the distance traveled by the escaping object.
This would lead to an extremely large and cumbersome differential
equation, and thus fairly unpractical to implement. 

\section{Application when the trajectory of the escaping object is other than
a straight line}

\begin{figure}
\centering\includegraphics[scale=0.5]{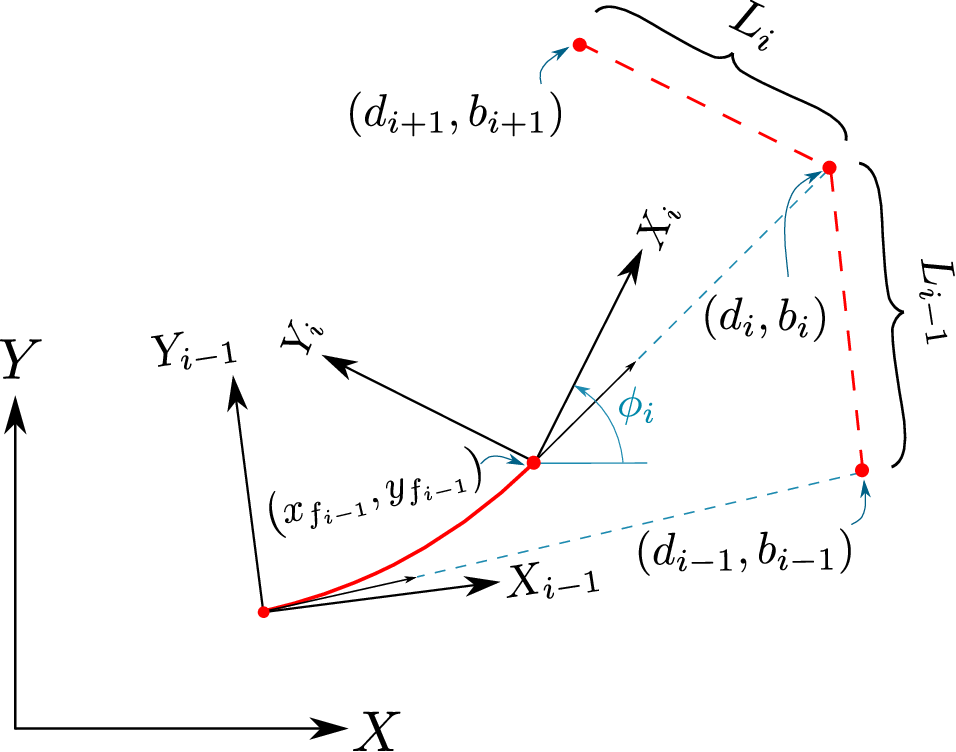}

\caption{The case when the escaping object follows an arbitrary path. In the
figure are shown two of the straight segments that approximate the
path of the escaping object, the coordinates systems associated with
each one of these straights segments are also shown.\protect\label{fig:case4}}
\end{figure}

The analysis developed so far have allowed us to obtain the trajectory
of the pursuit object when the escaping object moves in a straight
line. The most general case, which corresponds to an arbitrary path
can be solved if we consider this path as a set of straight segments.
If the trajectory of the escaping object is a smooth curve, then as
the length of these straight segments is reduced, which is to say
that its number increase, the approach to the trajectory of the escaping
object is becoming better. Let us consider the figure \ref{fig:case4}
in which two of the straight segments that approximate the actual
path of the escaping object are shown. For each one of these straight
segments there is a coordinate system associated to it. The coordinate
system is such that the $X_{i}$ axis is perpendicular to the i-th
segment and the $Y_{i}$ axis is parallel to it. The origin of this
i-th axis correspond to the final point of the pursuit object path
when follows the escaping object traveling along the (i-1)-th segment.
In this way, the problem of finding the trajectory of the pursuit
object when the escaping one follows an arbitrary path is reduced
to the repeated application of the procedure described in the preceding
sections.

The coordinates of the initial position of the escaping object, when
travels along the i-th segment, can be expressed in terms of the i-th
coordinate system as: 
\begin{equation}
\left\{ \begin{array}{c}
d_{i}^{i}\\
b_{i}^{i}
\end{array}\right\} =M_{\phi_{i}}^{-1}\cdot\left(\left\{ \begin{array}{c}
d_{i}\\
b_{i}
\end{array}\right\} -\left\{ \begin{array}{c}
x_{f_{i-1}}\\
y_{f_{i-1}}
\end{array}\right\} \right)
\end{equation}

where

\begin{equation}
M_{\phi_{i}}=\left[\begin{array}{cc}
\cos\left(\phi_{i}\right) & -\sin\left(\phi_{i}\right)\\
\sin\left(\phi_{i}\right) & \cos\left(\phi_{i}\right)
\end{array}\right]
\end{equation}
is the rotation matrix between the i-th and the principal X-Y coordinate
system, the vector $\left\langle x_{f_{i-1}},y_{f_{i-1}}\right\rangle ^{T}$is
the origin of the i-th coordinate system. All the super index indicates
in which coordinate system are measured the vector components\footnote{note that $\left\langle x_{f_{i-1}}^{i},y_{f_{i-1}}^{i}\right\rangle =\left\langle 0,0\right\rangle $},
if no super index is used then the term is assumed to be referred
to the principal X-Y coordinate system. The angle between the i-th
coordinate system and the principal X-Y system, $\phi_{i}$, can be
obtained as: 
\begin{equation}
\cos\left(\phi_{i}\right)=\frac{b_{i+1}-b_{i}}{\sqrt{\left(b_{i+1}-b_{i}\right)^{2}+\left(d_{i+1}-d_{i}\right)^{2}}}
\end{equation}

Finally we only need to know which are the initial velocities of the
pursuit and escaping object. These can be calculated as:
\begin{equation}
\left.\begin{array}{c}
\upsilon_{p0_{i}}=a_{p}t_{i-1}+\upsilon_{p0_{i-1}}\\
\upsilon_{e0_{i}}=a_{e}t_{i-1}+\upsilon_{e0_{i-1}}
\end{array}\right\} \label{eq:In_Vel}
\end{equation}
where $t_{i-1}=\tfrac{1}{a_{e}}\left(\sqrt{\upsilon_{e0_{i-1}}^{2}+2a_{e}L_{i-1}}-\upsilon_{e0_{i-1}}\right)$
is the total pursuit time when the escaping object travels along the
(i-1)-th segment, and $L_{i-1}$ is the length of the (i-1)-th straight
segment followed by the escaping object. With these information the
path that follows the pursuit object can be calculated using the procedures
described in the previous sections. 

\begin{figure}[H]
\centering\negmedspace{}\includegraphics[scale=0.6]{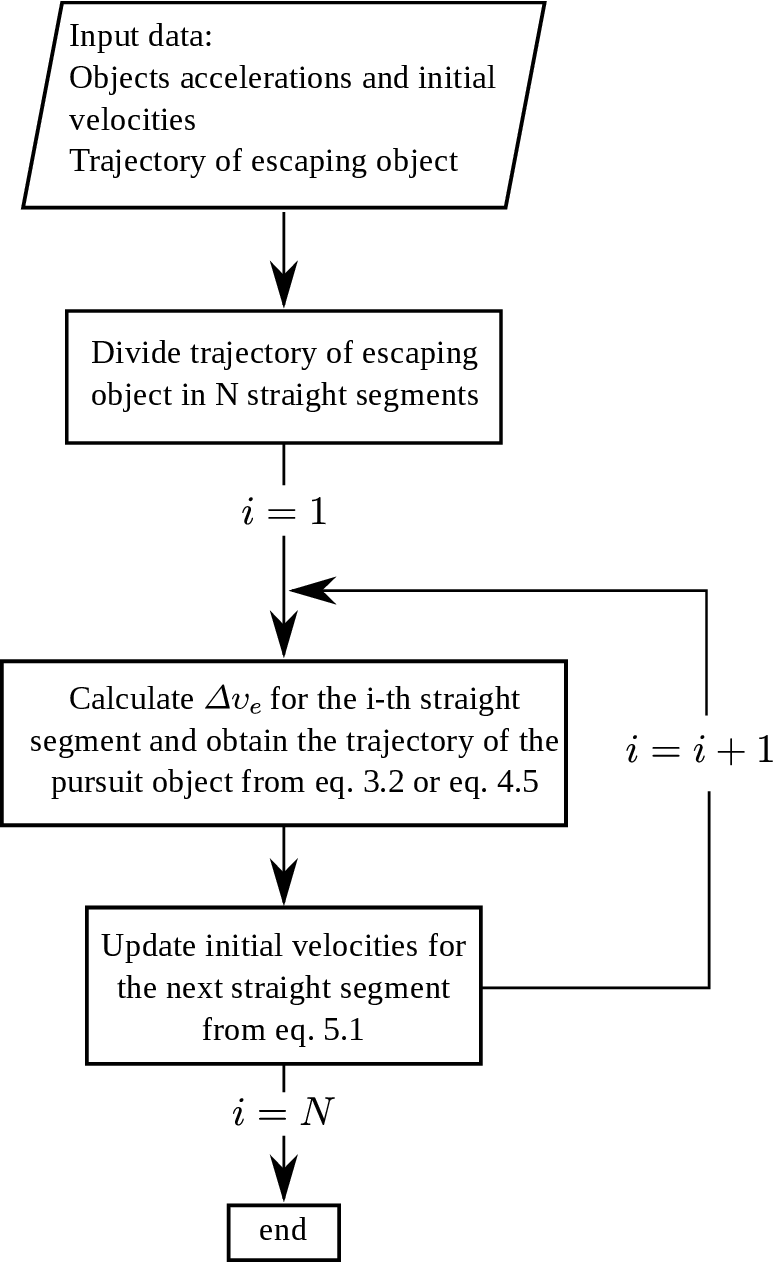}

\caption{Algorithm flow of the numerical procedure needed in order to obtain
the pursuit trajectory.\protect\label{fig:Flow}}
\end{figure}

The whole procedure described in this section can be summarized in
the flow chart shown in figure \ref{fig:Flow}, which involves three
basic steps. Firstly, the trajectory of the escaping object is divided
in $N$ straight segments. If this trajectory is indeed formed by
a set of straight segments then the value for $N$ can be directly
determined by their number. On the other hand if the trajectory is
a smooth curve then the value of $N$ should be decided in order to
obtain a good approximation of the escaping curve. In the second step
the value of $\Delta\upsilon_{e}$ is obtained, if this value is equal
to zero then equation \ref{eq:Case2-diffeq}, which has the exact
solution given by \ref{eq:TrayCaseI}, can be used to calculate the
corresponding portion of the pursuit curve, if $\Delta\upsilon_{e}\neq0$
then equation \ref{eq:Case3_EDO} should be used. Finally the values
of the initial velocities for the next segment are obtained from equation
\ref{eq:In_Vel}, and the process is repeated until the last straight
segment. 

\begin{figure}[H]
\centering\includegraphics[scale=0.6]{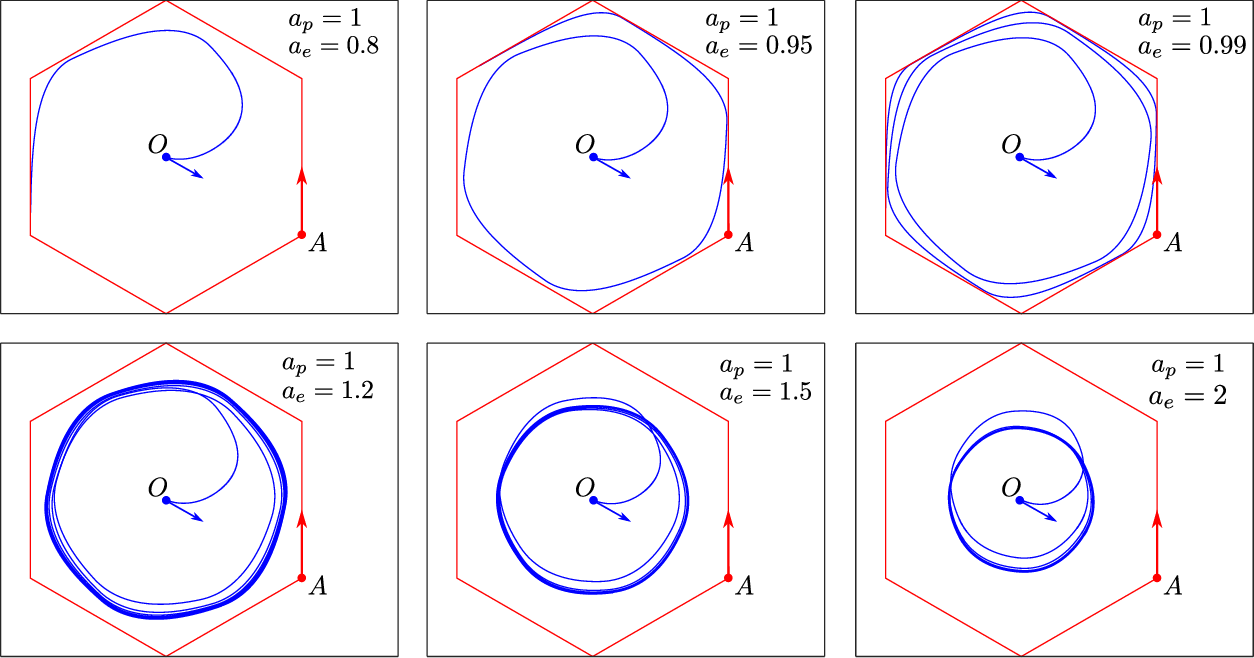}

\caption{The trajectory of the pursuit object when chase an object that moves
along an hexagonal path. The first three cases the pursuit object
manages to catch the escaping one ($a_{p}>a_{e}$), the last three
cases the escaping object is not reached by the pursuit one. In all
depicted cases $\upsilon_{e0}=4\upsilon_{p0}$.\protect\label{fig:Hex1}}
\end{figure}

\begin{figure}[H]
\centering\includegraphics[scale=0.6]{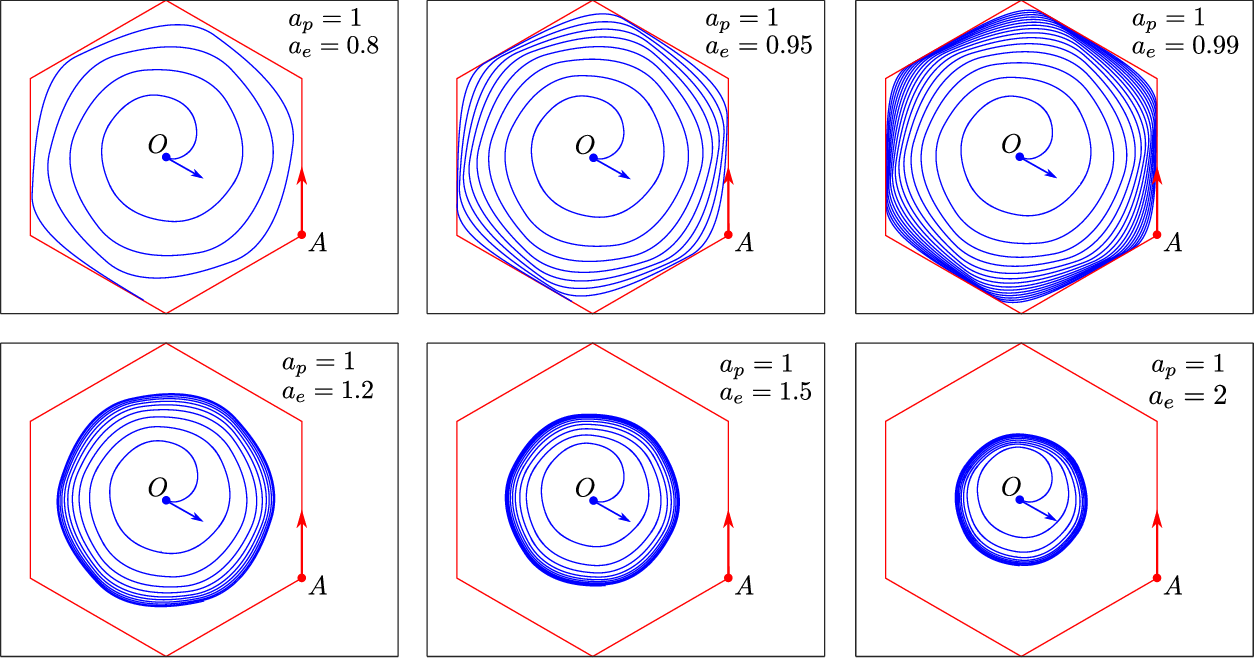}

\caption{The trajectory of the pursuit object when chase an object that moves
along an hexagonal path. The first three cases the pursuit object
manages to catch the escaping one ($a_{p}>a_{e}$), the last three
cases the escaping object is not reached by the pursuit one. In all
depicted cases $\upsilon_{e0}=10\upsilon_{p0}$.\protect\label{fig:Hex2}}
\end{figure}

As an example of the application of the procedure, let consider that
the escaping object follows a trajectory that has an hexagonal shape.
Using the Matlab solver ode45 the path of the pursuit object is calculated
for different initial velocities and acceleration ratios and are shown
in figures \ref{fig:Hex1} and \ref{fig:Hex2}.
 
In the first case where $a_{p}=1$ and $a_{e}=0.8$ the total pursuit
time is 75.257s, when $a_{p}=1$ and $a_{e}=0.99$ the total pursuit
time is 168.288 s. In the second case where $a_{p}=1$ and $a_{e}=0.8$
the total pursuit time reaches 176.044 s, and when $a_{p}=1$ and $a_{e}=0.99$
then the total pursuit time reaches 391.96 s. 

\begin{figure}[H]
\centering\includegraphics[width=0.85\textwidth]{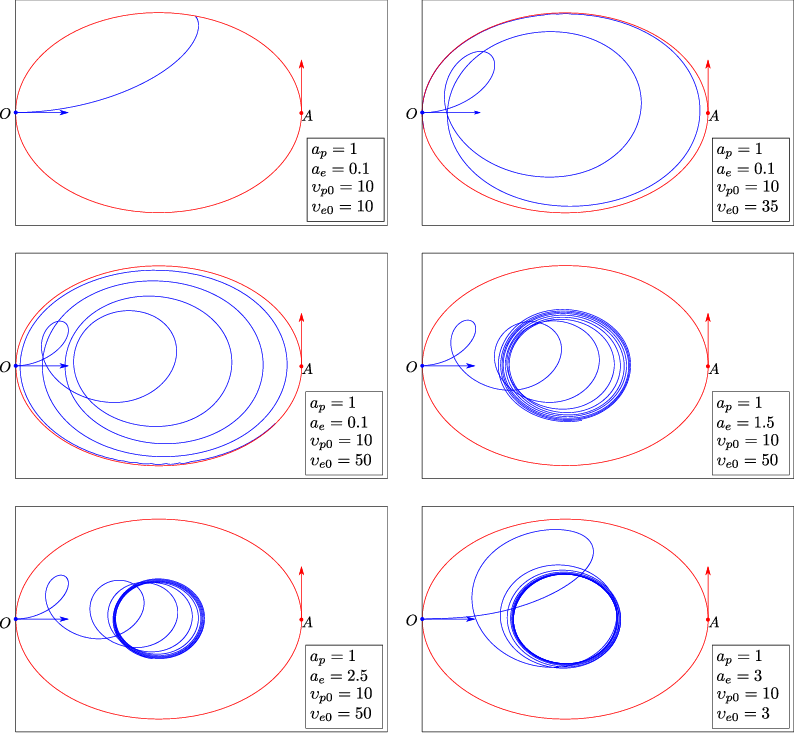}

\caption{The trajectory of the pursuit object when chase an object that moves
along an elliptical path. In the first three cases the pursuit object
manages to catch the escaping one ($a_{p}>a_{e}$), the last three
cases the escaping object is not reached by the pursuit one.\protect\label{fig:EllipPath}}
\end{figure}

As another example of the application of the procedure, in the figure
\ref{fig:EllipPath} is shown the trajectory of the pursuit object
when the escaping one moves around an elliptic trajectory. In this
case a total of 250 sections were used to approximate the elliptic
path of the escaping object.

The first three situations depicted in figure \ref{fig:EllipPath},
for which $a_{p}>a_{e}$, the chasing object succeeds in capturing
the escaping one. In the last three cases $a_{p}<a_{e}$ the capture
is not achieved, leaving the chasing object in a perpetual and futile
run. 

\section{Conclusions}

In the present study, the necessary analyzes have been developed to
obtain the trajectory of an object that is chasing another when both
have uniformly accelerated movements. The problem is first addressed
by solving the simplest case in which both objects start their motions
with no initial velocity. In this case the escape object is captured
if $a_{p}>a_{e}$, if this condition is met the total chasing time
is derived. Next, the case in which the initial speed of the escape
object is equal to $\alpha_{a}\upsilon_{p0}$ is studied, the capture
condition, and the chase time are also obtained. Its shown that this
last case can be used to solve the problem in which both objects have
arbitrary accelerations and initial velocities, as an alternative
to the direct approach that leads to an extremely large and cumbersome
differential equation. Finally, in order to obtain the trajectory of
the chasing object, when the escaping object moves following an arbitrary
trajectory, a procedure is presented in which the path of the escaping object is
approximated as a series of straight lines. This procedure
is used to solve two cases in which the escaping object moves following
a hexagonal and elliptical path. 

\bibliographystyle{plain}

\end{document}